\documentclass{aa}  

\usepackage{graphicx}
\usepackage{multirow}
\usepackage{amsfonts}
\usepackage{amssymb}
\usepackage[%
  breaklinks = true,
  colorlinks = true,
  urlcolor   = blue,
  citecolor  = blue,
  linkcolor  = blue,
]{hyperref}
\usepackage{soul}

\usepackage[normalem]{ulem}
\graphicspath{ {./figures/} }
\usepackage{txfonts}

\newcommand{\DtSTS}{\Delta t_{_{\rm STS}} }
\newcommand{\DtCFL}{\Delta t_{_{\rm CFL}}}
\newcommand{\DtMHDCFL}{\Delta t_{_{\rm MHD,CFL}} }
\newcommand{\DtADCFL}{\Delta t_{_{\rm AD,CFL}} }

\newcommand{\kb}{K_{\rm B}}
\newcommand{\etaamb}{\eta_{_{\rm amb}}}
\newcommand{\etahall}{\eta_{_{\rm Hall}}}
\newcommand{\etahallb}{\eta^{\ast}_{_{\rm Hall}}}
\newcommand{\etaambb}{\eta^{\ast}_{_{\rm amb}}}

\newcommand{\IRIS}{\textit{IRIS}}

\begin{document}

   \title{Ambipolar diffusion in the Bifrost code}
   \titlerunning{Ambipolar diffusion in the Bifrost code}

   \author{D. N\'obrega-Siverio\inst{1,2}
          \and
          J. Mart\'inez-Sykora\inst{3,4,1,2}
          \and
          F. Moreno-Insertis\inst{5,6}
          \and
          M. Carlsson\inst{1,2}
          }

   \institute{Rosseland Centre for Solar Physics, University of Oslo, PO Box 1029 Blindern, 0315 Oslo, Norway\\
   \email{desiveri@astro.uio.no}\\
             \and
             Institute of Theoretical Astrophysics, University of Oslo, PO Box 1029 Blindern, 0315 Oslo, Norway\\
             \and
              Bay Area Environmental Research Institute, NASA Research Park, Moffett Field, CA 94952, USA\\
              \and
              Lockheed Martin Solar and Astrophysics Laboratory, Palo Alto, CA 94304, USA\\
              \and 
              Instituto de Astrofisica de Canarias, Via Lactea, s/n, E-38205 La Laguna (Tenerife), Spain\\
              \and
              Department of Astrophysics, Universidad de La Laguna, E-38200 La Laguna (Tenerife), Spain
             }

   \date{Received February 24, 2020; accepted April 23, 2020}

%
%

 \abstract
  %
  %
   {Ambipolar diffusion is a physical
   mechanism related to the drift between charged
   and neutral particles in a partially ionized plasma that is key in many different astrophysical systems. However, understanding its effects is challenging due to basic uncertainties concerning relevant microphysical aspects  and the strong constraints it imposes on the numerical modeling.}
  %
  %
   {Our aim is to introduce a numerical tool that allows us to address complex problems involving ambipolar diffusion in which, additionally, departures from ionization equilibrium are important or high 
   resolution is needed. The primary application of this tool is for solar atmosphere calculations, but the methods and results presented here may also have a potential impact on other astrophysical systems.}
  %
  %
   {We have developed a new module for the stellar atmosphere Bifrost code that improves its computational capabilities of 
   the ambipolar diffusion term in the Generalized Ohm's Law. This module includes, among other things, collision terms adequate to processes in the coolest regions in the solar chromosphere. As a key feature of the module, we have implemented the Super Time-Stepping (STS) technique, that allows an important acceleration of the calculations. We have also introduced hyperdiffusion terms to guarantee the stability of the code.}
  %
  %
   {We show that to have an accurate value for the ambipolar diffusion coefficient in the solar atmosphere it is necessary to include as atomic elements in the equation of state not only hydrogen and helium but also the main electron donors like sodium, silicon and potassium. In addition, we establish a range of criteria to set up an automatic  selection of the free parameters of the STS method that guarantees the best performance, optimizing the stability and speed for the ambipolar diffusion calculations.
   We validate the STS implementation by comparison with a self-similar analytical solution.}
   {}


\maketitle

%
%
\section{Introduction}\label{sec:introduction}

When modeling astrophysical systems, the simplest magnetohydrodynamic (MHD) approximation is frequently used in which the plasma is considered 
as a single fluid with total coupling between its constituent microscopic species. This assumption is able to satisfactorily describe 
the physics of many phenomena in different astrophysical contexts; however, the approximation may no longer be valid when the plasma is partially ionized and ions and neutrals drift with respect to each other. 
This is the case for the 
interstellar medium \citep[e.g.,][]{Spitzer:1978,Zweibel:2002}, 
molecular clouds \citep[e.g.,][]{Zweibel:1983,Padoan:2000,Basu:2004,Crutcher:2012}, 
protoplanetary disks \citep[e.g.,][]{Wardle:1999,Salmeron:2008,Gressel:2015,Tomida:2015}
star formation \citep[e.g.,][]{Mestel:1956,Shu:1987,Kudoh:2008}
the solar chromosphere \citep[e.g.,][]{Goodman:2004, Zweibel:2011, Khomenko:2012,Martinez-Sykora2015,Zweibel:2015,Shelyag:2016}, among others.

It is possible to relax the MHD approximation to deal with partially ionized gases, considering the relative speed and 
associated friction between neutrals, ions and electrons, and still treating the plasma as a single fluid: the Generalized Ohm's Law 
(\citealp[see the seminal books by][]{Braginskii:1965,Mitchner:1973,Cowling:1976}). This way, 
the departure of the MHD approximation 
can be handled by just extending the induction and energy equations by adding the ambipolar diffusion term, which concerns the 
decoupling of neutral and charged components, and the Hall effect, which takes the drift velocities between ions and 
electrons into account. This extension has been applied in different codes by, for example,  \citealp[][among others,]{MacLow:1995,
Leake:2005rt,osullivan2007,Cheung:2012uq,Martinez-Sykora:2012uq,Masson:2012,Tomida:2015,Gonzalez-Morales:2018,Grassi:2019} and has 
been shown to be important to better understand the role of the ambipolar diffusion and Hall terms in astrophysics. However, the inclusion of partial ionization effects into advanced numerical codes confronts the modeler with different difficulties. 
On the one hand, the importance of the new effects sensitively depends on the microscopic constitution of the plasma, namely, on  
the abundances, the chemistry, the ionization degree and the collisions between different species. For instance, in the solar atmosphere, \cite{Martinez-Sykora:2012uq} showed that the approximation chosen to determine the values of collision cross sections and frequencies is crucial for ion-neutral interaction effects: there are significant discrepancies in the ambipolar diffusion coefficient depending on the assumption considered that lead to different results for the thermal properties, primarily in the chromosphere. In protostellar disc formation, \cite{Zhao:2016} found that reducing the number of very small grains enhances ambipolar diffusion. In molecular clouds, \cite{Grassi:2019} showed that cosmic rays can impact on the ionization level of the molecular gases, thus modifying the importance of the ambipolar diffusion. Those are a few examples of how the inclusion of proper physics is essential to obtain a realistic outcome when addressing partially ionized plasma.
On the other hand, the computations including partial ionization effects, even though being addressed from a single-fluid approach thus avoiding
the complexity of multifluid equations \citep[see, e.g.,][]{Leake:2012mf,Alvarez-Laguna:2016}, turn out to be very slow when solving them through explicit methods due to the strong constraints with respect to the timestep. 
According to the Courant-Friedrichs-Lewy (CFL) criterion 
\citep{Courant:1928uq}, the maximum timestep, $\DtCFL$, for the numerical solution of parabolic (e.g., diffusion) problems using explicit schemes 
decreases as the square of the spatial resolution $\Delta x$, that is, $\DtCFL \propto \Delta x^2/D$, where $D$ is the diffusion coefficient. Wherever high spatial resolution is required,  this quadratic dependence can strongly limit the 
calculation speed in comparison with non-diffusive MHD computations, whose $\Delta t$ is linearly dependent on $\Delta x$. As a consequence, high-resolution 
experiments of diffusion problems are virtually impossible  to perform explicitly. Different strategies have been carried out to alleviate this 
problem. For instance, \cite{Nakamura:2008,Li:2011, Masson:2012} use different thresholds to decrease the ambipolar term to avoid strongly restrictive timesteps when needed. Other authors like \cite{MacLow:1995, Mellon:2009, Leake:2006kx} adopt a sub-cycling method in which the induction equation is evolved separately from the rest of MHD equations when the timestep corresponding to the ambipolar diffusion is smaller than the dynamical timestep. An extension of this method is used by \cite{Martinez-Sykora:2012uq,Martinez-Sykora:2017sc,Martinez-Sykora:2017yo} to also consider 
sub-cycling the evolution of the energy equation because of the dissipation due to ambipolar diffusion. Another technique is the Super 
Time-Stepping \citep[STS;][]{Alexiades:1996}, which allows the restrictive CFL criterion to be relaxed to speed up the explicit calculation of
parabolic problems. This technique was shown to efficiently accelerate heat conduction calculations 
 \citep[see also][]{Meyer:2012,Iijima:2015},  and since then, it has been extensively used in ambipolar diffusion contexts \citep{choi2009,Commercon:2011,Tomida:2015,Gressel:2015,Gonzalez-Morales:2018}. The drawback is that the STS method has two free input
parameters, so it is necessary to carefully choose their values to not only optimize the performance but also to avoid the destabilization of the scheme which may lead to meaningless results 
\citep[for more details about numerical approaches in partially ionized systems, see the recent review by ][]{Ballester:2018}.

The purpose of this paper is to introduce a numerical tool that allows us to confront the numerical challenges
due to ambipolar diffusion in the solar atmosphere.  To that end, we have developed a new module in the Bifrost code \citep{Gudiksen:2011qy}, taking care, among other things, of the number 
of elements included in the calculations and their ionization state. Due to the numerical stiffness imposed by the ambipolar diffusion, its numerical implementation must be efficient to be able to calculate complex problems in which high resolution is mandatory.

The layout of this work is as follows. Section \ref{sec:gol} details the relevant equations of the Generalized Ohm's Law to establish the 
context for the subsequent parts of the paper. Section \ref{sec:collision} describes the implementation of the collision cross sections and 
frequencies necessary to compute the ambipolar diffusion coefficient. Section \ref{sec:ionization} addresses the computation of the ionization 
state for the ambipolar diffusion term when assuming local thermodynamic equilibrium (LTE), or nonequilibrium (NEQ) ionization and recombination 
of hydrogen and helium. In Section \ref{sec:STS}, we explain the STS method, together with its implementation in the Bifrost code. Section \ref{sec:hyperdiffusion} contains the recipes of the hyperdiffusion terms to guarantee stability for the code. Section \ref{sec:test} 
presents the validation test. Finally, 
Section \ref{sec:conclusions} summarizes the main conclusions of the present work.

%
%
\section{Generalized Ohm's Law}\label{sec:gol}
The Generalized Ohm's Law is basically a relation between the electric field and the electric current that makes it possible to overcome the difficulties of dealing with multifluid plasmas, such as
extremely high magnetic field mediated
wave speeds, a large number of equations, and stiff systems \citep[e.g.,][and references therein]{Ballester:2018}.
This is doable by using a one-fluid approximation that requires a high level of 
(but not infinite) coupling between neutrals and the charged species.
In a reference frame locally moving with a plasma element, 
it can be shown that this relation is given by 
\begin{equation}
      {\bf E'} = \eta {\bf J'}  -  
              \etaamb \frac{({\bf J'} \times {\bf B'})  \times {\bf B'}}{|\bf{B'}|^2}  + 
              \etahall \frac{({\bf J'} \times {\bf B'})}{|\bf{B'}|},
      \label{eq:gol}
\end{equation}
\noindent
where ${\bf E'}$ is the electric field, ${\bf B'}$ the magnetic field, and ${\bf J'}$ the current density 
all measured in that reference frame. The coefficient $\eta$ is the standard ohmic diffusion given by
\begin{equation}
      \eta  =  \frac{m_e (\nu_{en} + \nu_{ei}  )}{n_e q_{e}^2},
\label{eq:eta_ohm}
\end{equation}
\noindent
the ambipolar diffusion coefficient, $\etaamb$, is defined as 
\begin{equation}
      \etaamb = \frac{(\rho_N/\rho)^2|\bf{B'}|^2}{\Sigma_n \Sigma_i \rho_n \nu^{\ast}_{ni}} = 
      \etaambb|\bf{B'}|^2
\label{eq:eta_amb}
\end{equation}
\noindent
and the Hall coefficient $\etahall$ is given by
\begin{equation}
      \etahall = \frac{|\bf{B'}|}{q_e n_e}
      = \etahallb|\bf{B'}|
\label{eq:eta_hall}
\end{equation}
\noindent
where $q_e$ is the electron charge; $n_e$ the density number of electrons; $m_e$ the electron mass; 
$\rho_n$ is the density of the neutral element $n$; 
$\rho_N$ the total neutral mass density ($\rho_N=\Sigma_n \rho_n$); 
$\rho$ the total mass density; $\nu_{en}$ and $\nu_{ei}$ the 
collision frequency of electrons with neutrals and ions, respectively; and $\nu^{\ast}_{ni}$ is 
the reduced neutral-ion collision frequency 
that we will explain later in Section \ref{sec:collision}.
It is important to realize that to obtain Equation (\ref{eq:gol}), the relative velocity between charged species is considered negligible, and similarly between neutral species \citep[see][among others]{Zaqarashvili:2011, Khomenko:2014zr, Shelyag:2016}. 
This is an assumption that has been widely made in the literature to simplify the multifluid equations to a single fluid equation, and it implies that the collision times between charged species, on the one hand, and the neutral-neutral ones, on the other, must both be much smaller than the mixed charged-neutral collision times; and all of the foregoing must be much smaller than the macroscopic timescales. On the other hand, we would like to point out that the focus of this paper is on the ambipolar diffusion term and its new implementation in
the Bifrost code. The implementation of the Hall term has been presented in the papers by \cite{Martinez-Sykora:2012uq, Martinez-Sykora:2017yo}.

\subsection{The induction equation}
Going over now to the laboratory reference frame where the plasma element is moving with a velocity
${\bf u}$, the electric field, the electric current and the magnetic field are given by 
${\bf E} = {\bf E'} - {\bf u} \times {\bf B}$, ${\bf J} = {\bf J'}$, and ${\bf B} = {\bf B'}$, 
respectively. Using Faraday's induction equation and Equation (\ref{eq:gol}), one obtains a generalized 
induction equation, namely,
\begin{eqnarray}
      \frac{\partial {\bf B}}{\partial t} = \nabla \times \Big[    {\bf u} \times {\bf B} - \etahallb
      ({\bf J} \times {\bf B}) +  \etaambb({\bf J} \times {\bf B})  \times {\bf B}  - \eta {\bf J} \ \Big] 
\label{eq:induction1}
\end{eqnarray}
\noindent
This induction equation contains two additional terms in comparison to the one from the classic resistive MHD: 
a term proportional to ${\bf J} \times {\bf B}$, associated with the Hall effect; and another one proportional to $({\bf J} \times {\bf B}) \times {\bf B}$, associated with the ambipolar diffusion. Those terms cannot lead to changes in the 
magnetic topology and, consequently, neither produce magnetic reconnection. This can be seen 
if we define
\begin{equation}
 {\bf u}_{_{\rm Hall}} = - \etahallb {\bf J}
	\label{eq:u_hall}
\end{equation}
\noindent
and
\begin{equation}
      {\bf u}_{_{\rm amb}} = \etaambb({\bf J} \times {\bf B})
\label{eq:u_amb}
\end{equation}
\noindent
for the Hall and ambipolar terms, respectively. With those notations, 
\begin{eqnarray}
      \frac{\partial {\bf B}}{\partial t} = 
      \nabla \times \Big[ \big( {\bf u} +   {\bf u}_{_{\rm Hall}} + {\bf u}_{_{\rm amb}} \big )  
      \times  {\bf B}  - \eta  {\bf J} \ \Big] 
\label{eq:induction2}
\end{eqnarray}
which means that the magnetic field is no longer frozen into the plasma flow, but it is frozen 
into a pseudo-flow with speed ${\bf u} +   {\bf u}_{_{\rm Hall}} + {\bf u}_{_{\rm amb}}$. 
Even though those terms cannot lead to changes in the 
magnetic topology, they can significantly change the behavior of 
reconnection by, for example, a rapid thinning of the current sheet, or
an interplay with the plasmoid instability \citep[see, e.g.,][]{Huang:2011,Ni:2015}; when
$\eta = 0$, the topology is preserved and there is no magnetic reconnection.

\subsection{The energy equation}\label{sec:energy}
From basic electrodynamics, the power exerted by the electromagnetic field on the plasma is given by 
${\bf J'} \cdot {\bf E'}$. Using the Generalized Ohm's Law (Equation \ref{eq:gol}), we see 
that, in addition to the classic ohmic dissipation $\eta J^2$, there appears a new term that leads 
to an irreversible entropy increase, namely,
\begin{equation}
        Q_{_{\rm amb}} = \etaamb J_{\perp}^2,
\label{eq:q_amb}
\end{equation}
\noindent
where $J_{\perp}$ is the current component perpendicular to the magnetic field. This dissipation term 
is associated with the collisions between neutrals and ions and, hereafter, we refer to it as ambipolar 
diffusion heating. As a consequence, when taking ion-neutral interaction effects into account, a new entropy 
source has to be added to the energy equation as follows:
\begin{eqnarray}
      \frac{\partial e}{\partial t} = - \nabla  \cdot (e {\bf u})   -  P  \nabla  \cdot  {\bf u} + 
      \eta J^2 + \etaamb J_{\perp}^2 + Q_{_{\mathrm{rad}}} + Q_{_{\mathrm{Spitz}}}
\label{eq:energy}
\end{eqnarray}
\noindent
where $e$ is the internal energy per unit volume, $P$ the gas pressure, $Q_{_{\mathrm{rad}}}$ represents all 
the entropy sources due to radiation, and $Q_{_{\mathrm{Spitz}}}$ is the entropy source due to the thermal 
Spitzer conductivity.  The Hall term does not cause any energy dissipation since it is perpendicular to ${\bf J}$.

%
%
\section{Collision frequency and cross section}\label{sec:collision}
Collision frequency and cross section are important quantities when determining the rate at which electrons and the different 
ions and neutrals interact with each other. For this reason, we have to take the state-of-the-art models and measurements of the mentioned parameters into consideration .

\subsection{Collision frequency}\label{sec:collision_frequency}
The reduced neutral-ion collision frequency, $\nu^{\ast}_{ni}$, is given by
\begin{equation}
    \nu^{\ast}_{ni} = \frac{m_{ni}}{m_n} n_{_{i}} \sigma_{ni} \left( \frac{8 \kb T}{\pi m_{ni}}\right)^{1/2}
	\label{eq:nu_ni}
\end{equation}
\noindent
where $n_i$ the ion number density; $\kb$  the Boltzmann 
constant; $T$ the temperature; and $m_{ni} = m_n m_i/(m_n + m_i)$ the reduced mass of the neutral 
and ion species. In the solar atmosphere, the most frequent collisions
are those of the most abundant elements, H and He, with ions of other elements \citep{Khomenko:2014zr}. Therefore, we consider the following ion-neutral interactions: neutral hydrogen (H) and helium 
(He) atoms colliding with singly ionized ions of the 16 most important elements (either high abundance or low ionization potential) in the Sun and electrons, 
as well as neutral hydrogen molecules (H$_2$) colliding with protons (p) and electrons (e). 
Thus, the denominator of the ambipolar diffusion coefficient (Equation \ref{eq:eta_amb}) in our case is
\begin{equation}
\begin{split}
        \Sigma_n \Sigma_i \rho_n \nu^{\ast}_{ni}  
        = \rho_{\rm H} (\Sigma_i\nu^{\ast}_{\mathrm{H},i} + \nu^{\ast}_{\mathrm{H},e}) 
        + \rho_{\rm He} (\Sigma_i\nu^{\ast}_{\mathrm{He},i} + \nu^{\ast}_{\mathrm{He},e}) \\
        + \rho_{\rm H_2} ( \nu^{\ast}_{\mathrm{H_2},\mathrm{p}} + \nu^{\ast}_{\mathrm{H_2},\mathrm{e}} )
\end{split}
\label{eq:nu}
\end{equation}

\subsection{Cross section}\label{sec:cross_section}

\begin{figure}[ht]
\centering
\includegraphics[width=0.5\textwidth]{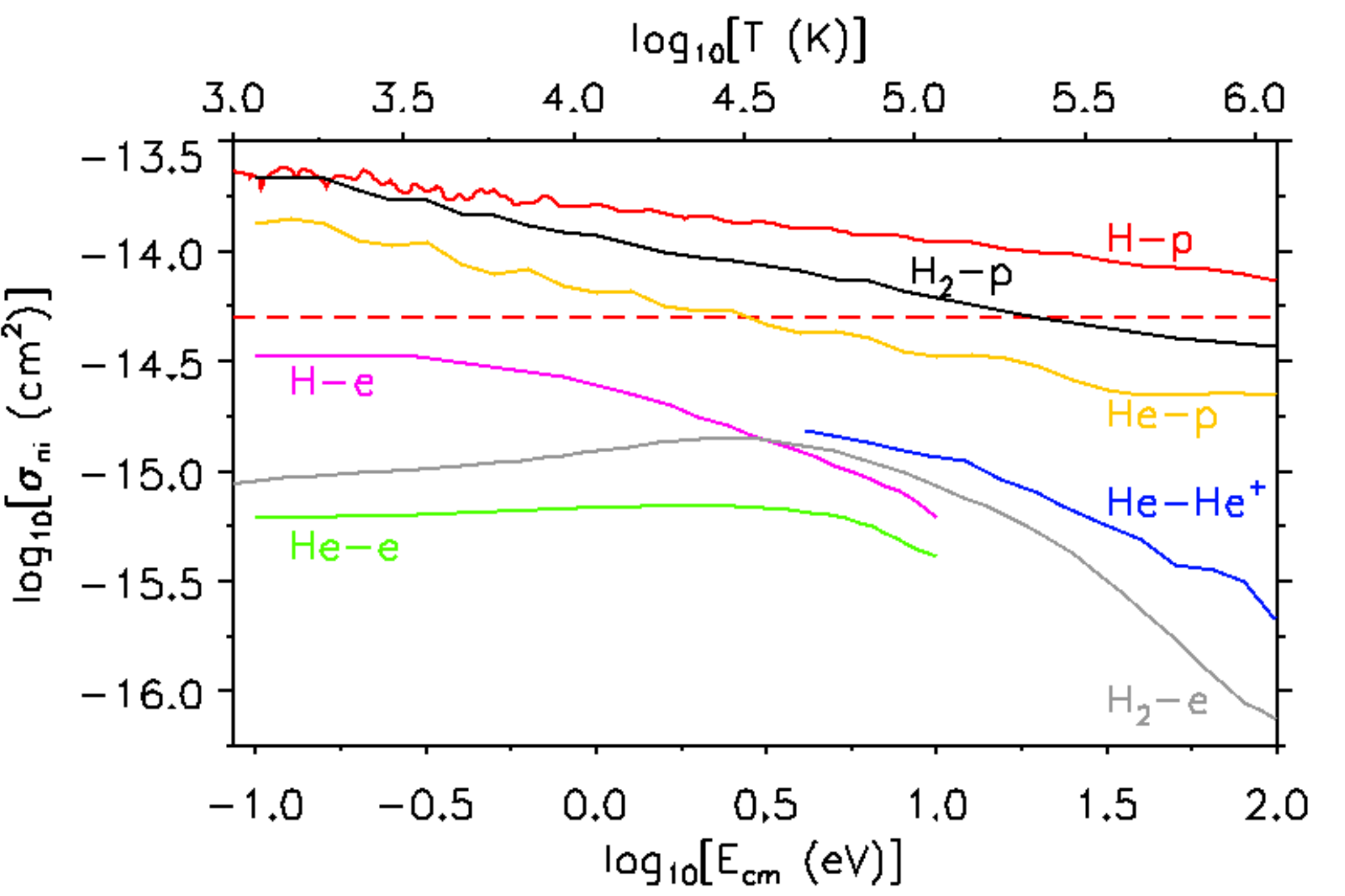}
\caption{Elastic scattering cross section, $\sigma_{ni}$, implemented in the Bifrost code to calculate $\etaamb$.  The different lines 
correspond to H--p collisions including temperature dependency and charge exchange (solid red line);
H--e (pink); 
He--p (yellow); 
He--He$^{+}$ (blue);
He--e (green); 
H$_2$--p (black);
and H$_2$--e (gray).
For comparison purposes, we have added the constant $\sigma_{ni}$ curve corresponding to H--p collisions
that is frequently used in the literature (dashed red line).}
\label{fig:01}
\end{figure}

The cross section for elastic scattering between a given neutral and a charged particle, 
$\sigma_{ni}$, is implemented through tables calculated on the basis of values given by different 
authors:

\begin{itemize}

    \item {\bf Neutral hydrogen atoms with protons (H--p).} From $10^{-4}$ to $10^{2}$~eV in the center of mass of the collision ($E_{\rm CM}$), the cross section values are based on quantum-mechanical 
    indistinguishability calculations, additionally including charge transfer 
    \cite[see][]{Krstic:1999,Glassgold2005,Vranjes:2013ve}\footnote{For larger values of the energy $E_{\rm CM}$, namely, $10^{2}$ to $10^{6}$~eV, consult the paper by \cite{schultz2008}}. 
    In Figure \ref{fig:01}, we 
    plot this cross section with a solid red curve in the range $E_{\rm CM} = [ {\sim}0.1, 100]$~eV, 
    which corresponds to typical solar temperatures $T~=~[10^3, {\sim}10^6]$~K 
    ($1$~eV~$= 11604$~K).
    For comparison purposes, we have also plotted, as a dashed red line, the constant cross section for collisions between neutral hydrogen atoms and protons used by other authors such as \cite{Khodachenko:2004vn,Leake:2006kx,Soler:2009,Khomenko:2012}, among others.
    For temperatures below $10^4$ K, the constant definition underestimates, by approximately an order of magnitude, the
    temperature-dependent cross section.
    
    \item {\bf Neutral hydrogen atoms with electrons (H--e).} The values are extracted from \cite{Vranjes:2013ve} 
    and references therein. In Figure \ref{fig:01}, this cross section is shown as a pink line from $0.1$ to $10$~eV.
    
    \item {\bf Neutral helium atoms with protons (He--p).} The elastic scattering cross sections are taken from \cite{Vranjes:2013ve} and shown in Figure \ref{fig:01} through a yellow curve.
    
    \item {\bf Neutral helium atoms with singly ionized helium (He--He$^{+}$)}. The values are obtained from the second chapter of the book by \cite{Franz:2009}. In Figure \ref{fig:01}, this cross section is shown in blue.
    
    \item {\bf Neutral helium atoms with electrons (He--e).} These values are extracted from \cite{Vranjes:2013ve} 
    and references therein. In Figure \ref{fig:01}, this cross section is plotted in 
    green from $0.1$ to $10$~eV.
    
    \item {\bf H$_2$ molecules with protons (H$_2$--p).} This cross section is based on the fully quantum-mechanical calculations by \cite{Krstic:1999mol} and is shown in Figure \ref{fig:01} with a black line.
    
    \item {\bf H$_2$ molecules with electrons (H$_2$--e).} Values of this elastic scattering cross section are 
    extracted from a data base compiled by \cite{Yoon:2008} and shown in 
    Figure \ref{fig:01} as a gray line.
    
    \item The collision cross sections for hydrogen (or helium)  and heavier elements are not well-known, 
    so we adopt the same assumption as made by \cite{Vranjes:2008uq}: we take the cross section between hydrogen (or helium) and protons multiplied by $m_i/m_{_{H}}$ (or $m_i/m_{_{He}}$), where $m_i$ is the mass of the heavier element.
    
\end{itemize}

%
%
\section{Ionization state}\label{sec:ionization}
The number of elements considered and their ionization state are 
also crucial to properly determine the ambipolar diffusion coefficient. 
For this reason, the module presented in this paper has been developed to compute $\etaamb$ in a consistent way with the 
different existing available equation-of-state (EOS) modules in the Bifrost code.
We have set up two methods to calculate $\etaamb$ according to the EOS used:
obtaining the ambipolar diffusion coefficient through precomputed tabulated values,  
which is only possible when using an EOS that assumes LTE (Section \ref{sec:LTE}); or
getting the ambipolar diffusion coefficient \textit{on the fly}, which is mandatory for 
NEQ ionization and recombination calculations (Section \ref{sec:NEQ}).

\subsection{LTE}\label{sec:LTE}
For LTE numerical experiments, Bifrost has an EOS based on tables generated by the 
\textit{Uppsala Opacity Package} \citep{Gustafsson:1975}, which computes instantaneous molecular dissociation equilibria and the LTE ionization balance 
of the 16 most important elements: elements with a high abundance (larger than 7 on a logarithmic scale) complemented with elements with an abundance down to 5 but with low ionization potential and therefore important as electron donors at low temperatures
\citep[see][for details]{Gudiksen:2011qy}. Table \ref{table:eos} shows the list of those 16 elements 
together with the  abundances, $A$, and atomic mass. Abundances were kept at the original values of
\citet{Gustafsson:1975} to ensure compatibility with simulations by \cite{Stein:2000}.

\begin{figure}[ht]
\centering
\includegraphics[width=0.49\textwidth]{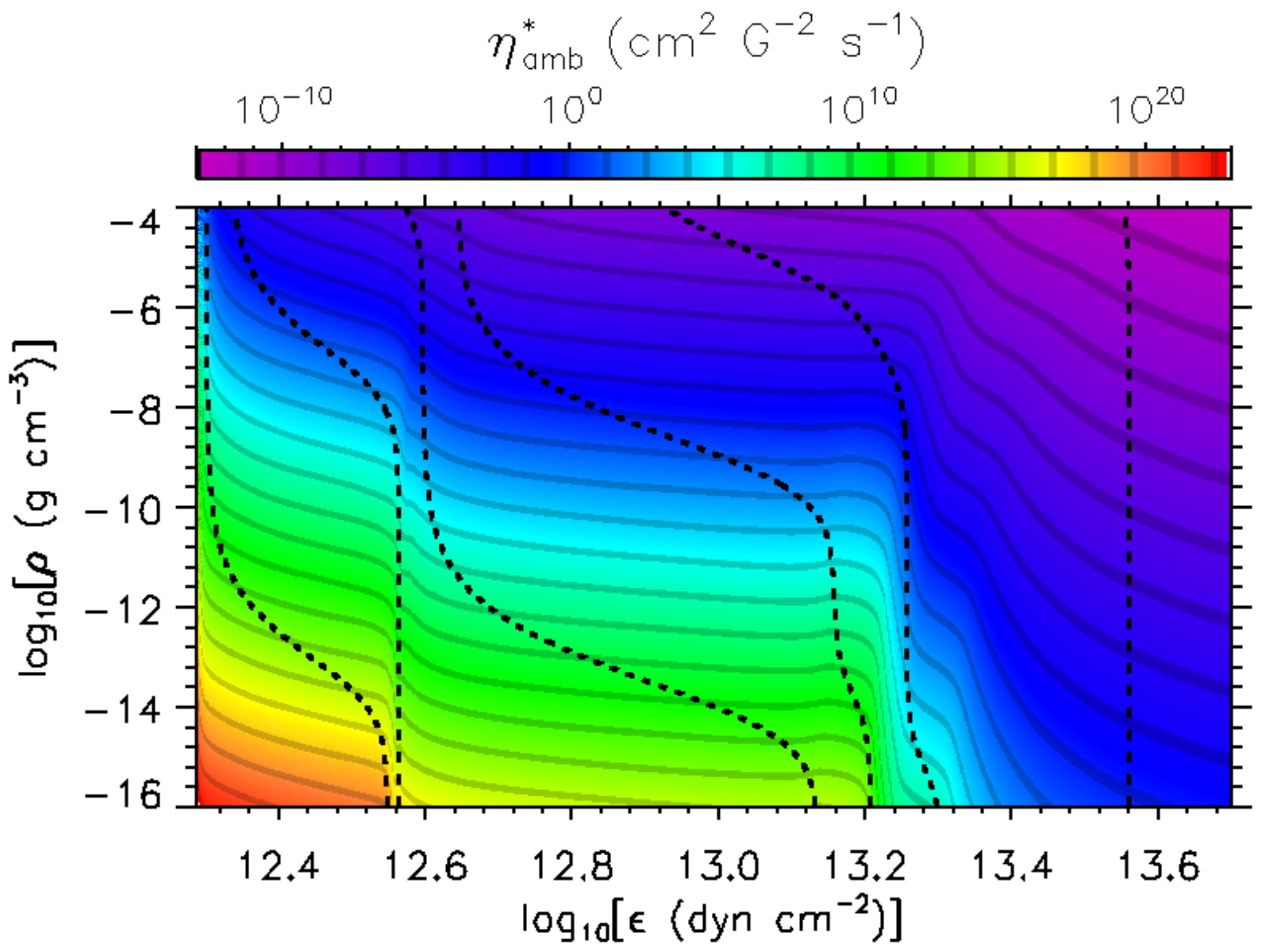}
\caption{Tabulated $\etaambb$ as a function of density, $\rho$, and internal energy per unit volume, $\epsilon$. 
The values for $\epsilon$ are shifted $5$~eV to obtain a positive
internal energy when having hydrogen molecules (see main text).
Temperature isocontours are superimposed as dashed lines for, from left to right, $3\times10^3$~K, $6\times10^3$~K, 
$10^4$~K, $2\times10^{4}$~K, and $10^5$~K.} 
\label{fig:02}
\end{figure}

In order to tabulate the ambipolar diffusion coefficient, we create a table for the $\etaambb$ coefficient, defined through Equation~\ref{eq:eta_amb}, which does not depend on the magnetic field, and can therefore be given just as a function of the density, $\rho$, and internal energy per unit volume, $\epsilon$. When executing numerical experiments, interpolations are carried out to get the 
ambipolar diffusion coefficient at the input energy and density. Figure \ref{fig:02} shows $\etaambb$ as a function of $\rho$ and $\epsilon$ from the aforementioned EOS table with temperature isocontours 
superimposed. The zero-point for the excitation/ionization/dissociation energy part of the internal energy is arbitrary. We have chosen this energy to be $5$~eV for a completely neutral gas with no molecules. This ensures a positive internal energy even when all hydrogen is bound in molecular form. 

\begin{figure*}[ht]
\centering
\includegraphics[width=\textwidth]{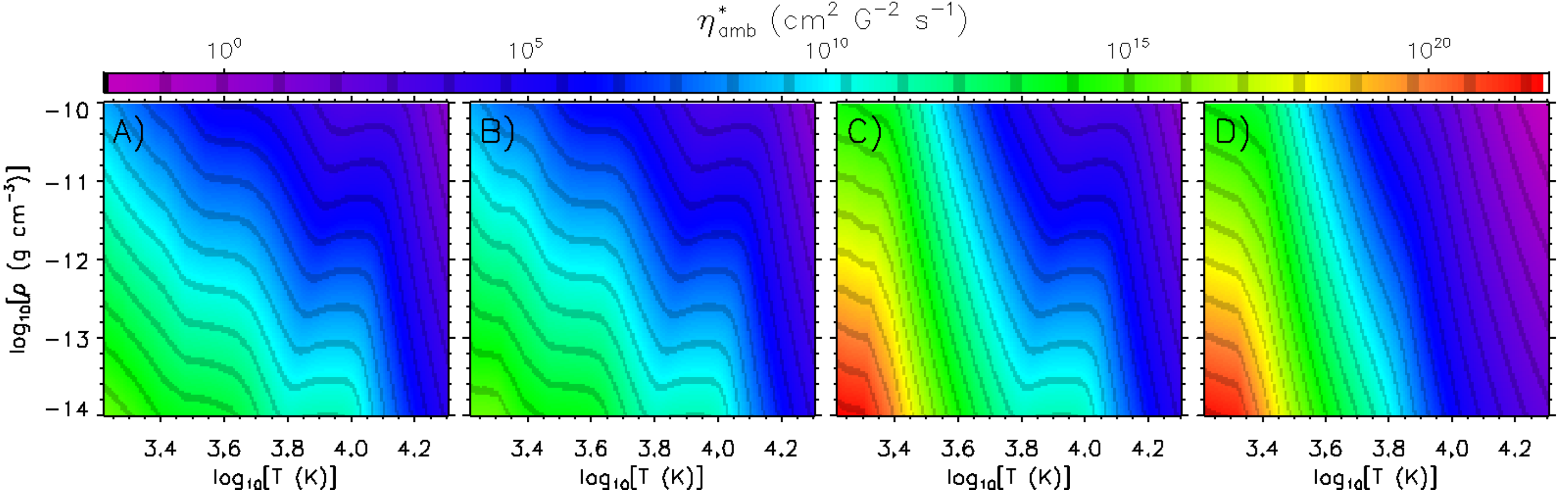}
\caption{Tabulated $\etaambb$ as a function of the density, $\rho$, and temperature, $T$, for different
EOS. \textit{Panel A}, the complete EOS of Bifrost. \textit{Panel B}, an EOS only considering atoms of H, He, Na, Si, Mg, K, and 
molecules of H$_2$. \textit{Panel C} an EOS just taking H, H$_2$, and He into account. \textit{Panel D} an EOS only including atomic H.} 
\label{fig:03}
\end{figure*}

We find that the ambipolar diffusion coefficient has a strong dependence on the number of elements
considered in the EOS. In order to show this important fact, in  Figure~\ref{fig:03} we compare  
$\etaambb$, as a function of $\rho$ and $T$, obtained using modified versions of the EOS of Bifrost. From left to right: A) the complete EOS of Bifrost; B) an EOS only considering atoms 
of H, He, Na, Si, Mg, K, and molecules of H$_2$; C) an EOS just taking H, H$_2$, and He into account; 
and D) an EOS only including atomic H.
Comparing panels A and D, we see that considering a pure hydrogen plasma leads to an overestimation of the role of the ambipolar diffusion by several orders of magnitude in the coolest temperatures. The reason is that as the temperature decreases and hydrogen becomes neutral, the total number of ions drops more quickly than when heavier elements are present, resulting in  larger values of the ambipolar diffusion coefficient. Assuming a pure hydrogen plasma also underestimates the ambipolar diffusion above $\log(T)\sim3.8$, in this case because,  as the temperature increases, 
hydrogen gets ionized leading to a lack of neutrals. Looking at panels A and C now, it is clear that introducing helium helps to get the right values of $\etaambb$ above $\log(T)\sim3.8$. Introducing molecules of H$_2$ slightly modifies the behavior in the range of temperature below $\log(T)\sim3.4$ (see the changes in the isocontours of the color scale); however, the values of $\etaambb$ are still overestimated in this regime due to the lack of the main electron donors. Panel B shows that we can start to get the right order of magnitude of the ambipolar diffusion coefficient once we introduce some of the main electron donors in the chromosphere, such as sodium, silicon, magnesium and potassium.

\begin{table*}
\caption{Abundances ($A$) \citep{Gustafsson:1975} and atomic mass ($Z$) for the
16 elements (El.) used in the EOS of Bifrost.}
\tiny 
\label{table:eos}      
\centering          
\begin{tabular}{|| c | c | c | c | c | c | c | c | c | c | c | c | c | c | c | c | c ||}     
\hline\hline
\rule{0pt}{2.5ex} 
El. & H & He & C & N & O & Ne & Na & Mg & Al & Si & S & K & Ca & Cr & Fe & Ni \\
\hline\hline
$A$ & 12.00 & 11.00 & 8.55 & 7.93 & 8.77 & 8.51 & 6.18 & 7.48 & 6.40 & 7.55 & 7.21 & 5.05 & 6.33 & 5.47 & 7.50 & 5.08 \\ 
\hline
$Z$ & 1.01 & 4.00 & 12.01 & 14.01 & 16.00 & 20.18 & 23.00 & 24.32 & 26.97 & 28.06 & 32.06 & 39.10 & 40.08 & 52.01 & 55.85 & 58.69 \\
\hline     
\end{tabular}
\end{table*}

\subsection{NEQ} \label{sec:NEQ}
In the previous section we show that the inclusion of elements heavier than hydrogen hugely impacts on the ambipolar diffusion coefficient calculation. This fact has been illustrated in the foregoing section through different LTE EOS; nonetheless, in the solar atmosphere, there are important departures from ionization equilibrium \citep[see the seminal papers by][]{Klein:1976,Klein:1978,Kneer:1980,Carlsson:1992,Carlsson:2002wl}, mainly in the chromosphere and transition region \citep[e.g.,][]{Bradshaw:2003,Bradshaw:2006,olluri:2013aa,olluri:2015,Martinez-Sykora:2016,Nobrega-Siverio:2018,Kerr:2019,Rutten:2019}. This means that it is not only important to include all the relevant elements, but also 
to get the ion and neutral number density for them through a nonequilibrium ionization calculation. This is moreover crucial
to compute the ambipolar diffusion coefficient due to its strong dependency on the number density of
ions and neutrals. In the following, the cases of nonequilibrium ionization of hydrogen and helium are commented separately.

\subsubsection{NEQ of hydrogen} \label{sec:NEQ:H}
The Bifrost code has a module available that calculates the ionization and recombination of atomic hydrogen and the formation and dissociation 
    of molecular hydrogen, H$_2$, under NEQ conditions 
    \citep[see][for details about this module]{Leenaarts:2007,Leenaarts:2011}.
    The ambipolar diffusion module presented in this paper can take the atomic and molecular hydrogen number densities from this NEQ module to compute
    the collision frequency (Equation \ref{eq:nu_ni}) and the ambipolar diffusion coefficient (Equation \ref{eq:eta_amb}), while the ionization state of the rest of the elements is assumed in LTE.
    Using the NEQ of hydrogen module together with the new ambipolar diffusion module presented here,  \cite{Nobrega-Siverio:2020} recently showed that
    the LTE assumption can lead to an important underestimation of the ionization fraction in
    magnetic flux emerging regions of up to 2-3 orders of magnitude. 
    This means that flux emergence experiments carried out under the LTE assumption significantly overestimate 
    the role of the ambipolar diffusion.

\subsubsection{NEQ of hydrogen and helium} \label{sec:NEQ:He}
The Bifrost code also has the capability of computing the NEQ ionization and recombination of helium by means of a module developed by \cite{Golding:2016}.
    The NEQ ionization and recombination of helium is important in the upper chromosphere,   where the helium ion fractions considerably depart from their equilibrium values. 
    This module always works together with the NEQ module of hydrogen, so when running ambipolar diffusion experiments, we take the atomic
    and molecular hydrogen, and atomic helium number densities from this module
    to calculate $\nu_{ni}$ and $\etaamb$, while the rest of elements are assumed in LTE.
    The combination of these modules largely impacts on the ambipolar diffusion distribution and changes the ambipolar heating in comparison to computations under the LTE assumption \citep{Martinez-Sykora:2020}, which can provide a completely new interpretation of the observations in the chromosphere \citep{Martinez-Sykora:2020alma}.

%
%
\section{Super Time-Stepping (STS) method}\label{sec:STS}
 
The Super Time-Stepping \citep[STS;][]{Alexiades:1996} is a technique that can be used to accelerate explicit parabolic calculations. It is based on the stability properties of the Chebyshev polynomials, which allow us to relax the CFL condition and take a larger timestep, $\DtSTS$. The method uses two input parameters, $n$, a positive integer, and $\nu \in (0,1)$, which is a damping factor. The timestep allowed by the STS method is then given by

\begin{equation}
	\DtSTS =
	\frac{n}{2 \sqrt{\nu}} \left[ \frac{ (1 + \sqrt{\nu})^{2n} -  
	(1 - \sqrt{\nu})^{2n}}{(1 + \sqrt{\nu})^{2n} + (1 - \sqrt{\nu})^{2n}}
          \right] \DtADCFL  \;,
\label{eq:sts1}
\end{equation}
\noindent
and is divided into $n$ sub-timesteps $\tau_i$ as follows:
\begin{equation}
	\DtSTS = \sum_{i=1}^{n} \tau_i
\end{equation}
\noindent
with
\begin{equation}
	\tau_i = \DtADCFL \left[ (\nu - 1) \cos{\left(  \frac{\pi (2i - 1)}{2n} \right)} + (\nu + 1) \right]^{-1}\;,
	\label{eq:sts2}
\end{equation}
\noindent
where $\DtADCFL$ is the timestep of the parabolic problem (in our case, ambipolar diffusion) given by the 
classic CFL condition. 
It is important to note that the intermediate values computed along the 
$n$ sub-timesteps have no approximating properties: it is only after the whole 
$\DtSTS$ has been reached that the results approximate the solution and, 
consequently, have a physical meaning.

The above expressions can be easily implemented to improve the calculation efficiency of ambipolar diffusion experiments; nevertheless, the drawback is that the STS is a first order scheme in time. In addition, the method has two free parameters, 
$n$ and $\nu$, and it is necessary to carefully choose their values to optimize the performance while keeping  numerical stability. The maximum ratio $\DtSTS/\DtCFL$ that can be reached for any given $n$ corresponds to the following limit:
\begin{equation}
	 \lim_{\nu \rightarrow 0} \frac{\DtSTS}{\DtADCFL} = 
	 \lim_{\nu \rightarrow 0}  \frac{n}{2 \sqrt{\nu}} \left[ \frac{ (1 + \sqrt{\nu})^{2n} -  
	 (1 - \sqrt{\nu})^{2n}}{(1 + \sqrt{\nu})^{2n} + (1 - \sqrt{\nu})^{2n}} \right] = n^2  
\label{eq:sts3}
\end{equation}
\noindent
In this limit, the CFL criterion would need $n^2$ steps to reach one $\DtSTS$, while 
the STS method requires just $n$, as explained above (Equation~\ref{eq:sts2}). Assuming 
a similar computing load for each step, whether in the STS or in the CFL-limited calculation, 
this implies that the STS method would be $n$ times faster than the simple CFL-limited one. 
However, it is necessary to impose a lower threshold for $\nu$, since $\nu=0$ is a stability 
limit, and choosing small values of $\nu$ can make the STS method very sensitive to round-off 
errors \citep{Alexiades:1996}. In the next subsection, we explain our choices 
made for $\nu$ and $n$.

\subsection{Implementation of the STS method in the Bifrost code}\label{sec:5.2.1}
In this section, we show the STS implementation in the Bifrost code and the 
choice of the $n$ and $\nu$ parameters. To that end, we use an operator splitting scheme, which
is a standard technique to advance in time the solution 
at each timestep separately for groups of terms that contain different physics, hence with different numerical requirements, while keeping codes modular.  
In our case, the induction and 
energy equations (Equations \ref{eq:induction1} and \ref{eq:energy}, respectively) 
are separated and solved over a timestep $\Delta t$ determined by the minimum
of the two following values: 1) the standard timestep given by the MHD and 
radiation terms considered in the Bifrost code 
\citep[see][for further details]{Gudiksen:2011qy}; and 2) the timestep imposed by 
the ambipolar and Hall term that appear in the generalized Ohm's law from Equation 
(\ref{eq:gol}). In particular, we take advantage of the operator splitting to
solve separately the ambipolar diffusion term using, when necessary, the STS
method. In order to determine whether the STS method can be efficiently applied, let us
assume a system in which 
\begin{equation}
       \DtMHDCFL = C\; \DtADCFL, \quad C \in \mathbb{R}^+.
\label{eq:c}
\end{equation}
\noindent
In this equation, $\DtMHDCFL $ is the timestep given by the CFL criterion for 
the standard MHD part, and $\DtADCFL$, the one for the ambipolar diffusion term. 
For any acceleration method, like the STS, to be of interest, it is obviously 
necessary that $ \DtMHDCFL > \DtADCFL$, that is, $C > 1$.  

\subsubsection{Choice of the free parameters $n$ and $\nu$}\label{sec:n_nu}
In the literature, even though STS has been broadly implemented in various codes and physical processes 
(see Section \ref{sec:introduction}), there is no thorough analysis of the choice of the 
$n$ and $\nu$ parameters, so the use of the STS method is not precisely specified. We have investigated several 
properties of these STS free parameters. In the following, we detail various aspects that should 
be considered:

\begin{enumerate} 
 \item $\DtSTS$ should not be larger than the timestep
   imposed by the general MHD and radiation terms ($\DtMHDCFL$), since the
   time advance of the system is limited by the minimum of all applicable
   timestep conditions. Consequently, 
 \begin{equation}
	\DtADCFL \leq \DtSTS \leq \DtMHDCFL \;.
	\label{eq:deltas1}
 \end{equation}
 \noindent
This condition can be rewritten using Equations (\ref{eq:sts1}) and (\ref{eq:c})
: 
    \begin{equation}
	1 \leq \frac{n}{2 \sqrt{\nu}} \left[ \frac{ (1 + \sqrt{\nu})^{2n} -
            (1 - \sqrt{\nu})^{2n}}{(1 + \sqrt{\nu})^{2n} + (1 -
            \sqrt{\nu})^{2n}} \right] \leq C 
	\label{eq:deltas3}
  \end{equation}
\noindent
with $\nu \in (0,1)$ and $C > 1$. The above is the minimum requirement 
to apply the STS method. For $n=1$ the function at the center
of the inequality is below 1 for all $\nu > 0$, so, since $n$ is an integer, Equation 
(\ref{eq:deltas3}) implies $n\geq2$.

  \item In addition, we tighten the lower bound of Equations (\ref{eq:deltas1}) and (\ref{eq:deltas3})
   as follows: for the STS method to be computationally advantageous, its $n$
   substeps must permit a greater advance in time than $n$ times the timestep 
   $\DtADCFL$. Thus, in  Equation (\ref{eq:deltas1})
   we set $n \DtADCFL < \DtSTS$, so the lower bound of 
   Equation (\ref{eq:deltas3}) becomes
      \begin{equation}\label{eq:def_offnnu}
	n < \frac{n}{2 \sqrt{\nu}} \left[ \frac{ (1 + \sqrt{\nu})^{2n} -
            (1 - \sqrt{\nu})^{2n}}{(1 + \sqrt{\nu})^{2n} + (1 -
            \sqrt{\nu})^{2n}} \right] \equiv f(n,\nu)
  \end{equation}
\noindent
For simplicity, we have introduced the symbol $f(n,\nu)$ for the term on the right of the above expression. It can be shown that the constraint (\ref{eq:def_offnnu}) imposes a maximum value for $\nu$. To prove it, in Figure \ref{figure_04} we have plotted $f(n,\nu)$ against $n$ for different values of $\nu$ (dashed lines). 
In the image, the solid red curve is $f(n,\nu)=n^2$, which corresponds to the limit of stability $(\nu = 0)$.
 The solid black curve is simply the straight $f(n,\nu)=n$ that allows us to discern
which combinations of $n$ and $\nu$ verify the above inequality. As seen in the figure,
$f(n,\nu)=n$ roughly coincides with the curve $\nu = 0.25$. In fact, through 
a limit analysis of $f(n,\nu)/n$,
  \begin{equation}
	 \lim_{n \rightarrow \infty}  \frac{1}{2 \sqrt{\nu}} \left[ \frac{ (1 + \sqrt{\nu})^{2n} -
            (1 - \sqrt{\nu})^{2n}}{(1 + \sqrt{\nu})^{2n} + (1 -
            \sqrt{\nu})^{2n}} \right] = \frac{1}{2\sqrt{\nu}},
  \end{equation}
  \noindent
   we find that the constraint $f(n,\nu)/n=1$ gives us an asymptotic limit of $\nu$, namely, $\nu  = 0.25$: 
    the maximum value of $\nu$ that results from the optimization constraint (\ref{eq:def_offnnu}). In the code, to avoid working with asymptotic limits, we use as the 
   upper limit of $\nu$, the one obtained for the lowest possible $n$, i.e., $(n=2)$, namely, $\nu < \sqrt{5} - 2$. 
  
\begin{figure}[ht]
\centering
\includegraphics[width=0.45\textwidth]{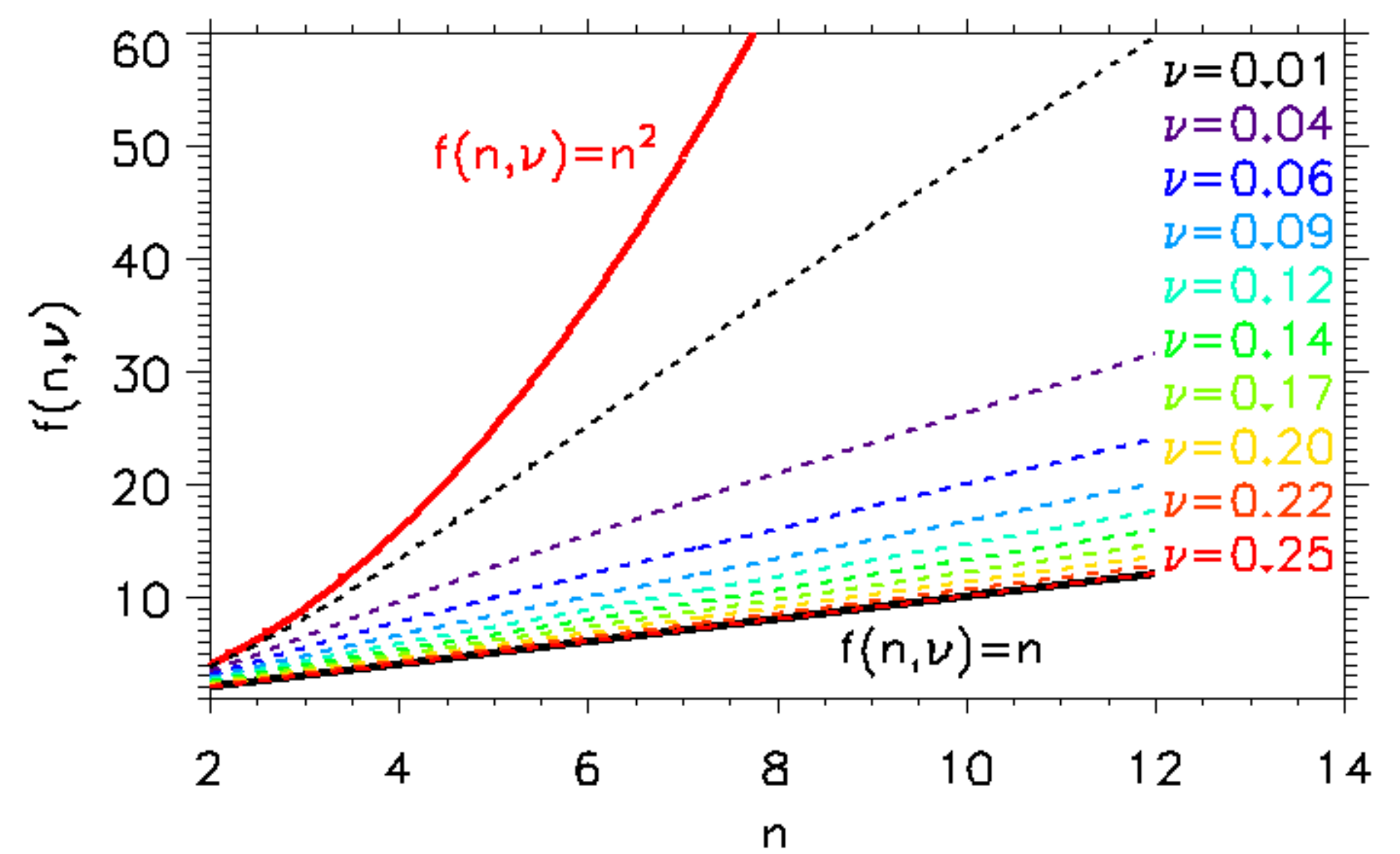}
\caption{Plot showing $f(n,\nu)$ versus $n$ for different values of $\nu$ (dashed curves). Solid lines represent
$f(n,\nu)=n^2$ (red) and $f(n,\nu)=n$ (black).} \label{figure_04}
\protect
\end{figure}

\item There is also a strong constraint concerning the use of large $n$. In principle, it would be mathematically
possible to have any value of $n$; nonetheless, numerically, large values of $n$ can be problematic. This is 
because, during the $n$ sub-cycling shown in Equation (\ref{eq:sts2}), the inner values obtained within the loop 
do not have any physical meaning, as already mentioned, and can reach very large values that compromise the accuracy of the calculations due to the limited precision. There are three ways to alleviate this problem: (a) increasing from single to double precision, which allows calculations with larger values of $n$ but increases the memory requirements; (b) 
introducing an elaborate normalization, which helps to avoid Infinity or !NAN values but increases the number of 
operations and complicates its implementation; (c) imposing a strict maximum of $n$ according to the experience. 
The latter is the solution we have adopted and requires some testing
to discern the valid $n$ that will keep the simulations stable. We find that $n$ starts to be problematic above 12, approximately. In order to ensure 
greatest stability, we have decided to impose as a maximum $n=10$. This, together with the previous
considerations, gives us the two following ranges we must fulfill: $0 < \nu < \sqrt{5} - 2$ and $2 \leq n \leq 10$.
In Figure \ref{figure_05}, the valid parameter-space derived from those ranges is shown within the three solid red lines of the image.

\begin{figure}[ht]
\centering
\includegraphics[width=0.45\textwidth]{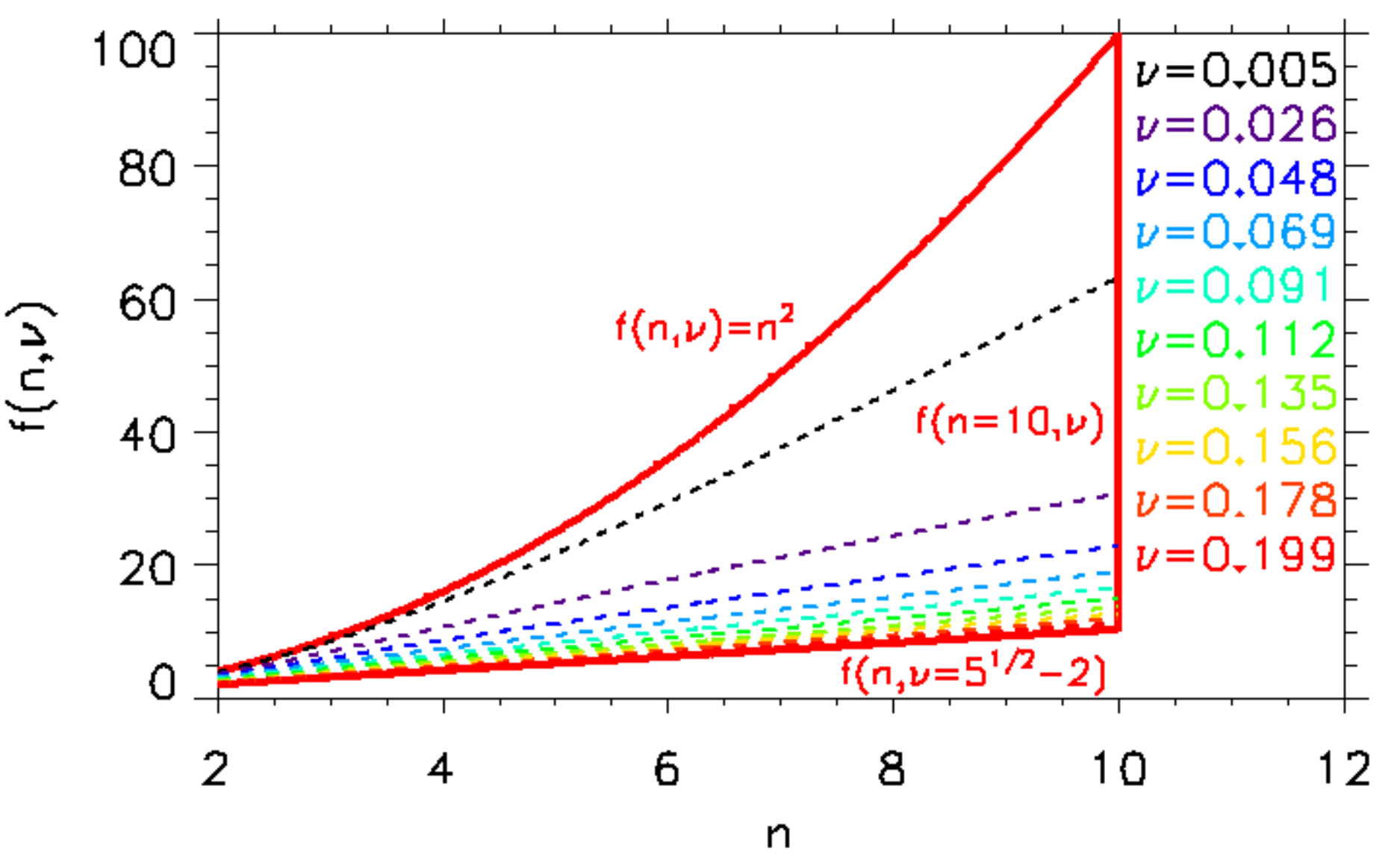}
\caption{Plot showing $f(n,\nu)$ versus $n$ for different values of $\nu$ (dashed curves). Solid lines represent
the boundaries of the parameter-space considering the constraints described in the text related to 
optimization and stability, namely, $f(n,\nu)=n^2$, $f(n=10,\nu)$ and, $f(n,\nu=\sqrt{5}-2)$.} 
\label{figure_05}
\protect
\end{figure}

\item The final step to determine the optimum combination of $n$ and $\nu$ can be
taken now. The ideal situation would be to reach $\Delta t_{_{MHD,CFL}}$ in the minimum
number of stable $n$ steps without compromising the accuracy of the solution. To that end, we impose,
\begin{equation}
\DtSTS = \Delta t_{_{MHD,CFL}}
\end{equation}
\noindent
which implies the following relationship between $n$  and $\nu$:
\begin{equation}
  f(n,\nu) = \frac{n}{2 \sqrt{\nu}} \left[ \frac{ (1 + \sqrt{\nu})^{2n} -  (1 - \sqrt{\nu})^{2n}}{(1 + \sqrt{\nu})^{2n} + (1 - \sqrt{\nu})^{2n}}   \right] = C
 \label{eq:ligature}
  \end{equation}
  \noindent
We could obtain the minimum number of substeps $n$ in the limit $\nu \rightarrow 0$, so, from Equation (\ref{eq:sts3}) and considering that $n$ has to be an integer, $n=\mathrm{int}(\sqrt{C})$. 
However, since $\nu = 0$ corresponds to the limit of stability, one should consider $n > \mathrm{int}(\sqrt{C})$. 
The first option would be to use $n = \sqrt{C} + 1$; nevertheless, 
the associated $\nu$ could still be small to affect our calculations 
by round-off errors as pointed out by \cite{Alexiades:1996}. For this reason, in our STS calculations we impose
\begin{equation}
 n = \mathrm{int}(\sqrt{C}) + 2 \quad\hbox{with}\quad C > 4
  \end{equation}
\noindent
For the cases when $2 \leq C \leq 4$, instead of STS, we apply the subcycling method described 
by \cite{Martinez-Sykora:2017yo}, that is, only the induction and energy equation are 
evolved separately from the rest of MHD equations; when $1 \leq C \leq 2$, we follow the CFL criterion.

\end{enumerate}

\begin{figure}[ht]
\centering
\includegraphics[width=0.45\textwidth]{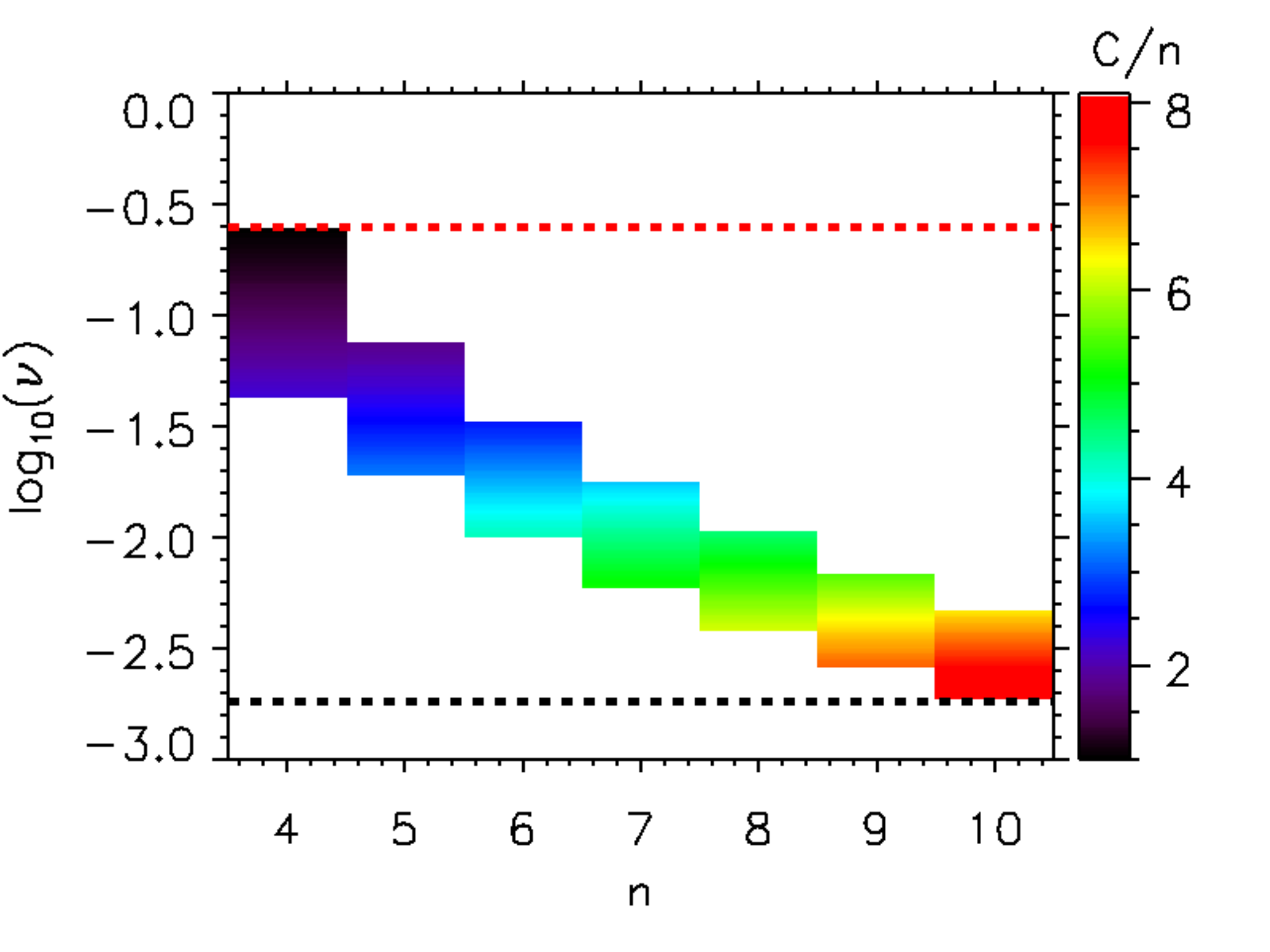}
\caption{2D map of $C/n$ showing as colored patches the combinations of $n$ and $\nu$ that verify Equation (\ref{eq:ligature}) for $C >4$. In the image, the horizontal dashed lines mark the values
$\nu = \sqrt{5} - 2$ (red) and the value of $\nu$ for the maximum $n=10$ ($\nu \approx 0.00185$, black)} \label{fig:06}
\protect
\end{figure}

Having determined $n$, we can see that, to apply the STS method, for each $n$ there is a limited range of values of $\nu$ that 
satisfies Equation (\ref{eq:ligature}) for $C >4$. This is illustrated in Figure \ref{fig:06} through a 2D 
map of $C/n$ showing the range of $\nu$ that satisfy that condition for each $n$ as a colored patch. The corresponding values of $C/n$ are given in the colorbar.
In the image, we have overplotted a red horizontal line to delimit $\nu = \sqrt{5} - 2$, and a black one that 
delineates the minimum value of $\nu$ for the maximum $n=10$ ($\nu \approx 0.00185$). We see that with this implementation, we can speed up the calculations up to a factor 8. Using the criteria just indicated for the choice of the $\nu$ and $n$ parameters, we can 
get the best performance, optimizing the speed for the ambipolar diffusion calculations in the Bifrost code.

%
%
\section{Hyperdiffusion}\label{sec:hyperdiffusion}
In order to maintain the stability of the code, we have implemented hyperdiffusion terms related
to the ambipolar diffusion. The diffusive operator for the induction equation follows the idea
originally described by \cite{mhdcodeboris} and later implemented in the Bifrost code \citep{Gudiksen:2011qy}
and in its preexisting non-STS Generalized Ohm's Law module \citep{Martinez-Sykora:2017yo}.
In the following, for compactness, we only describe the hyperdiffusion operator for the induction 
equation in the $x$ direction. Equivalent explanations can be given for the other two directions.

\begin{equation}
    \begin{split}
    \frac{\partial B_x}{\partial t} = ... 
    &+ \frac{\partial}{\partial z}\left[ \frac{\Delta x}{2}
    \left(\nu^{(1)}_x + \nu^{(2)}_x + \nu^{(3)}_x \right)Q\left(\frac{\partial B_z}{\partial x} \right)J_y 
    \right] \\
    &+ \frac{\partial}{\partial z}\left[  \frac{\Delta z}{2}
    \left(\nu^{(1)}_z + \nu^{(2)}_z + \nu^{(3)}_z \right)Q\left(\frac{\partial B_x}{\partial z} \right)J_y 
    \right] \\
    &- \frac{\partial}{\partial y}\left[  \frac{\Delta x}{2}
    \left(\nu^{(1)}_x + \nu^{(2)}_x + \nu^{(3)}_x \right)Q\left(\frac{\partial B_y}{\partial x} \right)J_z 
    \right] \\
    &- \frac{\partial}{\partial y}\left[  \frac{\Delta y}{2}
    \left(\nu^{(1)}_y + \nu^{(2)}_y + \nu^{(3)}_y \right)Q\left(\frac{\partial B_x}{\partial y} \right)J_z
    \right]
    \end{split}
\end{equation}
\noindent
where $\Delta x$, $\Delta y$ and $\Delta z$ are the grid spacing in the $x$, $y$ and $z$ direction, respectively;
$Q$ is a positive definite quenching factor \citep[see][]{mhdcodeboris,Gudiksen:2011qy,Martinez-Sykora:2017yo};
and $\nu^{(1)}_j$, $\nu^{(2)}_j$ and $\nu^{(3)}_j$
are the three hyperdiffusion terms in the $x$, $y$, or $z$ direction
that we describe in the following.

\begin{itemize}
    \item $\nu^{(1)}_j$ is a hyperdiffusion term
    proportional to the ambipolar diffusion coefficient that damps waves with the shortest wavelengths, that is, the largest wavenumber: the Nyquist wavenumber. The term is as follows
    \begin{equation}
          \nu^{(1)}_j  = \nu_1 \etaamb  \frac{\pi}{\Delta j}
    \end{equation}
    where $\nu_1$ is a dimensionless coefficient of order $10^{-2}$.
    \item $\nu^{(2)}_j$ is a hyperdiffusion term related to the advective transport of magnetic field attached to the charged species, in fact, more precisely, to the electrons. It is defined as
    \begin{equation}
        \nu^{(2)}_j = \nu_2  | u_{_{\mathrm{amb}}, j} |
    \end{equation}
    where $\nu_2$ is a dimensionless coefficient of order $10^{-2}$. The Bifrost code already has a hyperdiffusion term because of the advection of the magnetic field with the velocity $\vec{u}$, and for that
    reason here we only need to consider the extra 
    part due to ambipolar diffusion. 

    \item $\nu^{(3)}_j$ is a hyperdiffusion term related to magnetic shocks.
    The magnetic field is only influenced by the perpendicular velocity field,
    so it is possible to adopt a hyperdiffusion term that depends
    on the convergence of the velocity field. As in the previous term, Bifrost 
    has already implemented a term for $\vec{u_{\perp}}$, so here we only have 
    to implement one for the ambipolar diffusion velocity. Since this
    velocity field is already perpendicular, the $\nu^{(3)}_j$ term is as
    follows:
    \begin{equation}
        \nu^{(3)}_j = \nu_3 \Delta j  | \nabla \cdot \vec{u_{_{\mathrm{amb}}}} |_j
    \end{equation}
    where $\nabla$ is a first order finite difference operator and $\nu_3$
    a dimensionless coefficient of order $10^{-2}$.
    
\end{itemize}

%
%
\section{Validation test}~\label{sec:test}
After programming the STS following the parameter choice described in Section \ref{sec:STS} and explaining the hyperdiffusion terms
in Section \ref{sec:hyperdiffusion}, we need to verify that 
the STS is correctly implemented in the Bifrost code. To that end, we have proceeded with a test 
that has the advantage that allows us to verify the method in 2.5D MHD scenarios for different 
initial conditions. This is possible thanks to the existence of a self-similar analytical solution 
\citep[see][]{Pattle:1959}.
In the following, we describe this analytical solution and the goodness of our STS implementation.

\subsection{Self-similar solution for ambipolar diffusion problems}
The test we intend to carry out is based on the induction equation when only taking 
the ambipolar term into account, in other words,  the pure ambipolar diffusion case. If we also assume that the ambipolar coefficient 
$\etaambb$ is homogeneous, the induction equation (Equation \ref{eq:induction1}) becomes
\begin{equation}
\frac{\partial {\bf B}}{\partial t} =   \etaambb \nabla \times \left[
   ({\bf J \times B}) \times {\bf B} \right].
\label{eq:pureambipolar}
\end{equation}
\noindent
In a 2D domain with polar coordinates, this equation becomes a non-linear diffusion equation when 
considering an axisymmetric function for ${\bf B}$. \citeauthor{Pattle:1959}'s self-similar solution is of the form:
\begin{equation}
      B_y (r, t) \longrightarrow \left(\frac{\Phi}{4 \pi \etaambb t}  \right)^{1/3} \left[ 1 - \frac{r^2}{R^2 (t)}  \right]^{1/2},
\label{eq:selfsimilar4}
\end{equation} 
\noindent
for $r<R(t)$, and $0$ otherwise; where
\begin{equation}
      R^2 (t) \longrightarrow  \frac{3}{2} \left[ 4\etaambb t \left(\frac{\Phi}{\pi}  \right)^2 \right]^{1/3},
\label{eq:selfsimilar4b}
\end{equation} 
$\Phi$ is the magnetic flux, $y$ is the symmetry axis, and $r^2= (x-x_0)^2 + (z-z_0)^2$.
From those expressions, it is clear that $B_y \propto t^{-1/3}$ and  $R \propto t^{1/6}$.
The problem has the additional feature that any single-lobed initial condition
with total flux $\Phi$ is expected to tend, asymptotically in time, to the shape given in Equations 
(\ref{eq:selfsimilar4}) and (\ref{eq:selfsimilar4b}) \citep{Zeldovich_book1967}.  
This is a robust multidimensional test to check whether the STS method was properly implemented in the code.

\subsection{Numerical test}
\begin{figure}[ht]
\centering
\includegraphics[width=0.5\textwidth]{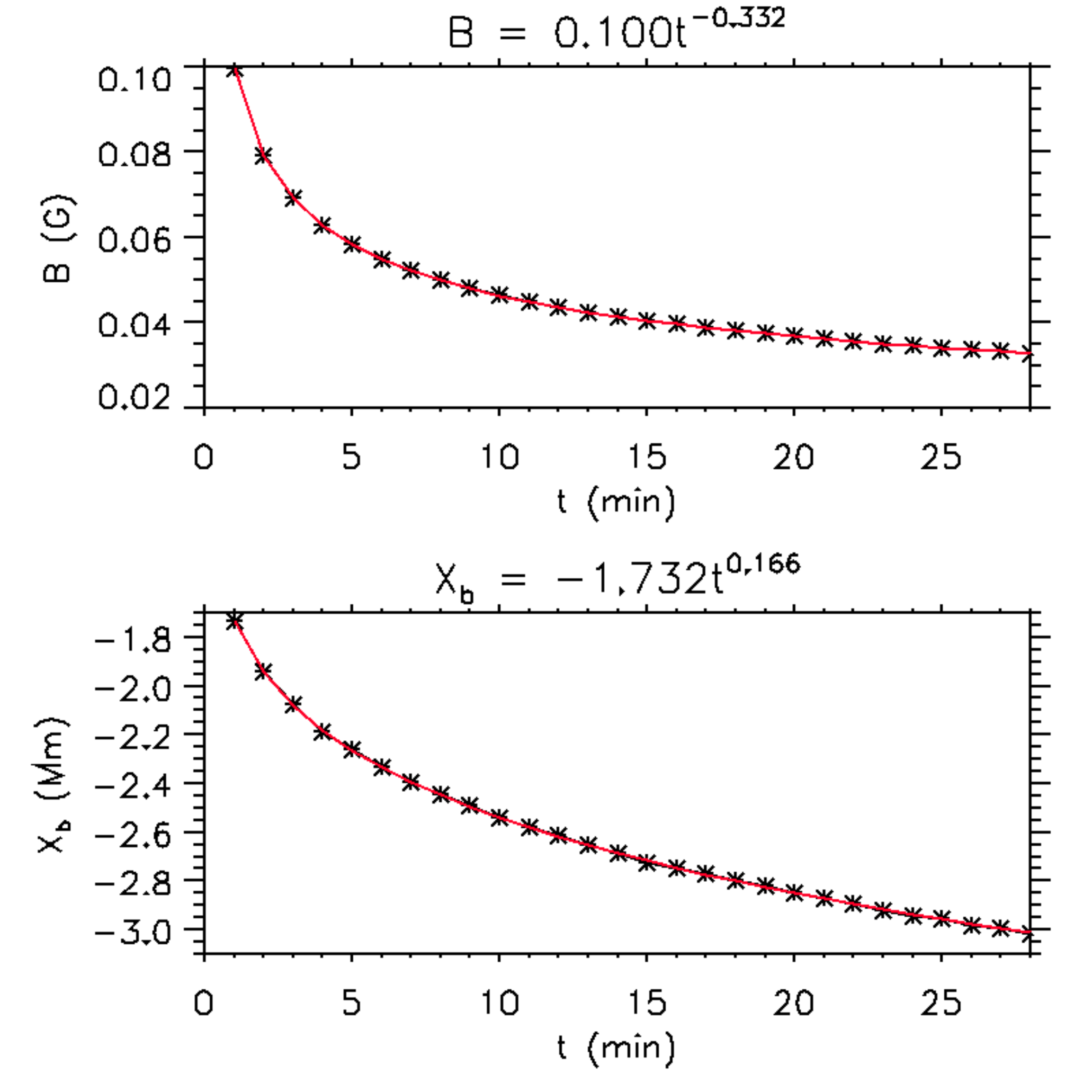}
\caption{Results of the STS validation test for the self-similar solution. Top panel: evolution of the maximum
of the magnetic field with time (black asterisks) and power-law fit (red line). Bottom panel: evolution of $X_b$ with time
(black asterisks) and power-law fit (red line). Above each panel is shown the resulting fit formula.} 
\label{fig:07}
\protect
\end{figure}
In order to test the STS method with that self-similar solution, we create a 2.5D snapshot 
of $1024\times1024$ points. The physical domain spans from $-5.0 \leq x \leq 7.5$ Mm 
to $-5.0 \leq z \leq 7.5$ Mm, with a numerical resolution of $\Delta x = \Delta z = 12$ km. 
In that domain,  we define as initial condition the following axisymmetric function for the magnetic field
\begin{equation}
  B_{y0} (x,z) = B_0 e^{ - \left[ (x-x_0)^2 + (z-z_0)^2 \right]/w^2}
\label{eq:selfsimilar4c}
\end{equation}  
where $B_0 = 0.8$ G, $x_0= 0$ Mm, $z_0= 0$ Mm, and $w= 0.5$ Mm. The rest of parameters are 
representative of the chromosphere: the initial
internal energy per unit volume is $e_0 = 4\times10^{-3}$ erg cm$^{-3}$; the initial density 
$\rho_0 = 10^{-9}$ g cm$^{-3}$; and the ambipolar diffusion coefficient is
set to a constant, namely, $\etaambb= 10^{16}$~cm$^{2}$~s$^{-1}$~G$^{-2}$.
This initial condition does indeed converge to the self-similar solution with the same integrated magnetic flux after an initial transient phase. The results for the self-similar stage are shown in Figure \ref{fig:07}. In the image, the top panel illustrates the evolution of the maximum of the magnetic field with time through black asterisks. The power-law fit to that curve is
\begin{equation}
  B_{y} = 0.100 \thinspace t^{-0.332},
\label{eq:selfsimilar4d}
\end{equation}  
and it is shown as a red curve in the figure. The exponent in (\ref{eq:selfsimilar4d}) agrees quite well with the predicted value, $-1/3$,  shown in Equation (\ref{eq:selfsimilar4}). On the other hand, the bottom panel contains the evolution of $X_b$ (one of the Cartesian components of $R$)
 with time (black asterisks) and its power-law fit fit (red). The power-law fit yields
\begin{equation}
  X_{b} = -1.732 \thinspace t^{0.166},
\label{eq:selfsimilar4e}
\end{equation}  
which again agrees with the expected value $1/6$ from Equation (\ref{eq:selfsimilar4b}). This tests indicates 
that our STS implementation is correct.
In addition, to ascertain the speed-up factor that we gain by means of the STS, 
we have run the same test with the same number of CPUs under the CFL criterion and also using the subcycling method described by \cite{Martinez-Sykora:2017yo}. The results show that for this test, STS can lead to a speed-up factor of 8 and 4 with respect to the CFL and subcycling method, respectively.

%
%
\section{Conclusions}\label{sec:conclusions}

In this paper, we have described the implementation of a new module in the Bifrost code 
to efficiently address ambipolar diffusion problems in the solar atmosphere
with the state-of-the-art microphysics, pursuing a fourfold purpose:

\begin{itemize}
    \item We have reviewed the existing literature to implement the most accurate 
    collision cross sections and frequencies, which are crucial for proper calculation 
    of the ambipolar diffusion coefficient. For instance, we have considered
    the temperature-dependent cross section of the collisions between hydrogen and protons
    \citep{Vranjes:2013ve}, which can be, roughly, one order of magnitude larger than the
    constant value usually taken in the literature 
    \citep[e.g.,][]{Khodachenko:2004vn,Leake:2006kx,Soler:2009}
    for temperatures below $10^4$ K.
    In addition, we have included the cross sections of collisions with hydrogen molecules that may be important in the coolest regions of the Sun, such as sunspot umbrae, where estimations from observations reveal that there is a 
    substantial H$_2$ molecule formation present \citep{Jaeggli:2012}.
    
    \item Through an analysis of the elements included in the EOS, we have shown that in the solar atmosphere, it is necessary to include heavier elements than hydrogen, primarily helium and the most important electron donors, to properly estimate the role of the ambipolar diffusion. In fact, considering a pure hydrogen plasma can lead to an overestimation of the ambipolar diffusion coefficient by several orders of magnitude.
    
    \item Since the correct determination of the ionization fraction is also relevant for the ambipolar diffusion calculation, we have also made this new module compatible with previous Bifrost modules that compute the NEQ ionization and recombination of hydrogen and helium as well as the H$_2$ formation under NEQ conditions \citep{Leenaarts:2011,Golding:2016}. This way, we have opened a new avenue to address partial ionization effects together with departures of the ionization state from the LTE in the solar atmosphere. In fact, NEQ ionization and recombination effects of hydrogen have been shown to be key for the understanding of the ambipolar diffusion in magnetic flux emergence processes in the Sun \citep{Nobrega-Siverio:2020}. In addition, \cite{Martinez-Sykora:2020}, combining the module presented in this paper together with the two aforementioned NEQ modules of hydrogen and helium, have shown the important variations in the thermodynamics in the chromosphere and transition region with respect to calculations carried out assuming LTE.
    
    \item We have also implemented the STS method in the Bifrost code, carrying out a thorough analysis of the choice of the two free parameters of the STS method to obtain the best performance. This choice allows us to speed up single-precision computations including the ambipolar diffusion term up to a factor 8 in comparison with the same calculation run under the CFL criterion. In addition, we have described the hyperdiffusion terms that maintain  the stability of the code when ambipolar diffusion is taken into account.
    
\end{itemize}

The numerical techniques presented in this paper may facilitate hints to address ambipolar diffusion 
problems in other astrophysical systems.

%
%
\begin{acknowledgements}
This research is supported by the Research Council of Norway through its
Centres of Excellence scheme, project number 262622, and through grants of
computing time from the Programme for Supercomputing. It is also supported by 
the Spanish Ministry of Science, Innovation and Universities through
projects AYA2014-55078-P and PGC2018-095832-B-I00, as well as through the
Synergy Grant number 810218 (ERC-2018-SyG) of the European Research Council; 
by NASA through 
grants NNX17AD33G, 80NSSC18K1285 and contract NNG09FA40C (\IRIS) and by the NSF grant AST1714955.
The authors thankfully acknowledge the computer
resources provided at the Pleiades cluster through the computing projects s1061, and s2053 from the High End Computing (HEC) division of NASA.
In addition, this study has been discussed within the activities of team 399
``Studying magnetic-field-regulated heating in the solar chromosphere’’ at the
International Space Science Institute (ISSI) in Switzerland.
\end{acknowledgements}

%
%
\bibliographystyle{aa}
\bibliography{collectionbib}

\end{document}